\documentclass[prl,aps,showpacs,floatfix,nofootinbib,twocolumn,letterpaper,reprint]{revtex4-1}\usepackage{wrapfig}
\usepackage{color,graphicx}
\usepackage{dcolumn}
\usepackage{bm}
\usepackage{verbatim}
\usepackage{multirow}
\newcommand{\mc}{\multicolumn}
\usepackage{amsmath}

\usepackage{epsfig}
\usepackage{amssymb}
\usepackage{amsmath}
\usepackage{amsfonts}
\usepackage{bm}
\usepackage{braket}

\newcommand{\TrA}{\text{Tr$_\mathcal{A}$ }}
\makeatletter

\newcommand{\Rmnum}[1]{\expandafter\@slowromancap\romannumeral #1@}

\newcommand{\mn}{-}

\newcommand{\be}{\begin{equation}}
\newcommand{\ee}{\end{equation}}
\newcommand{\ba}{\begin{eqnarray}}
\newcommand{\ea}{\end{eqnarray}}

\makeatother

\newcommand{\ssec}[1]{\emph{#1}.---}

\begin{document}

\title{Extracting spectra in the shell model Monte Carlo method using imaginary-time correlation matrices}

\author{Y. Alhassid,$^1$ M. Bonett-Matiz,$^2$ C.N. Gilbreth,$^3$ and S. Vartak}

\affiliation{Center for Theoretical Physics, Sloane Physics Laboratory, Yale University, New Haven, Connecticut 06520, USA\\
$^2$Physics Department, University of Bridgeport, Bridgeport, Connecticut 06511, USA\\
$^{3}$Quantinuum, Broomfield, Colorado 80021, USA
}

\begin{abstract}
Conventional diagonalization methods to calculate nuclear energy levels in the framework of the configuration-interaction (CI) shell model approach are prohibited in very large model spaces.  The shell model Monte Carlo (SMMC) is a powerful technique for calculating thermal and ground-state observables of nuclei in very large model spaces, but it is challenging to extract nuclear spectra in this approach. We present a novel method to extract low-lying energy levels for given values of a set of good quantum numbers such as spin and parity.  The method is based on imaginary-time one-body density correlation matrices that satisfy asymptotically a generalized eigenvalue problem.  We validate the method in a light nucleus that allows comparison with exact diagonalization results of the CI shell model Hamiltonian.  The method is applicable to other finite-size quantum many-body systems that can be described within a CI shell model approach.
\end{abstract}


\maketitle

\ssec{Introduction} Standard diagonalization methods in the framework configuration-interaction (CI) shell model~\cite{Brown1988,Caurier2005}  have been successfully used to extract the spectra of light and mid-mass nuclei. However, their applications to heavy open-shell nuclei are prohibited by the combinatorial increase of the dimensionality of the many-particle model space with number of valence orbitals and/or nucleons.   This difficulty has been overcome in part by the auxiliary-field quantum Monte Carlo (AFMC) approach~\cite{Johnson1992,Lang1993,Alhassid1994,Koonin1997,Alhassid2001,Alhassid2017}, also known in nuclear physics as the shell-model Monte Carlo (SMMC) method.  SMMC has enabled CI shell model calculations of thermal and ground-state observables in model spaces that are many orders of magnitude larger than those that can be addressed by conventional CI shell model techniques.  The method has been particularly useful in calculating statistical properties of nuclei~\cite{Nakada1997, Alhassid1999, Alhassid2007, Alhassid2008, Bonett2013, Alhassid2014,Ozen2013,Alhassid2015}.  However, it is a major challenge to extract spectroscopic information about individual excited levels.

In principle, spectral information of many-body systems can also be extracted by taking the Fourier transform of real-time response functions. However, in quantum Monte Carlo methods it is only possible to calculate imaginary-time response functions.   Here we introduce a method to extract energy levels in SMMC by computing imaginary-time correlation matrices (ITCM) of one-body densities for given values of good quantum numbers.  These matrices are shown to satisfy a generalized eigenvalue problem (GEVP) at zero temperature. The generalized eigenvalue (GEV) solutions are then used to extract
excitation energies.  The ITCM method enables us to calculate low-lying excited energy levels for given values of the good quantum numbers such as  spin and parity.  A GEVP has been used in lattice QCD to study excited states of baryons~\cite{Luscher1990,Beane2011,Beane2015,Edwards2013}.

We validate the ITCM method by applying it to a light $sd$-shell nucleus ($^{20}$Ne), for which the spectrum can be calculated exactly using conventional diagonalization methods~\cite{Johnson2018}. 

\ssec{SMMC}
The SMMC method is based on the representation of the many-body Gibbs operator $e^{-\beta \hat H}$ at inverse temperature  $\beta=T^{-1}$ as a coherent superposition of one-body propagators describing non-interacting nucleons moving in external time-dependent auxiliary fields $\sigma=\sigma(\tau)$. In this representation, known as the Hubbard-Stratonovich transformation~\cite{Hubbard1959,Stratonovich1957}, the Gibbs ensemble has the
form 
\be 
e^{-\beta\hat{H}}=\int\mathcal{D}[\sigma] G_\sigma\hat{U}_\sigma \;,
\ee 
where $\hat{H}$ is the many-body Hamiltonian (which includes a one-body term and a two-body residual interaction), $\mathcal{D}[\sigma]$ is the metric of the path integral, $G_\sigma$ is a Gaussian weight, and $\hat{U}_\sigma$ is a one-body propagator in imaginary-time $\tau$ between $\tau=0$ and $\tau=\beta$. 

We define for a $\sigma$-dependent quantity, $X_\sigma$, the following expectation value
\begin{equation}
\overline{X}_\sigma=\frac
{\int\mathcal{D}[\sigma]  W_\sigma\Phi_\sigma X_\sigma}
{\int\mathcal{D}[\sigma]  W_\sigma \Phi_\sigma}\,,\label{CanDef}
\end{equation}
where ${W_\sigma=G_\sigma |\TrA\hat{U}_\sigma|}$ is a positive definite weight function and 
$\Phi_\sigma=\TrA\hat{U}_\sigma/ |\TrA\hat{U}_\sigma|$ is the Monte Carlo sign for a given configuration of the auxiliary fields $\sigma$.  Here $\TrA$ denotes the trace for a fixed number of particles ${\mathcal A}$ calculated using particle-number projection~\cite{Ormand1994,Alhassid1999}. We project on both proton number $Z$ and neutron number $N$, so ${\mathcal A}$ refers to $(Z,N)$.

The thermal expectation values of an observable $\hat O$ in the canonical ensemble for a nucleus with $\mathcal{A}$ nucleons is then given by  ${\langle\hat{O}\rangle = \overline{\langle\hat{O}\rangle}_\sigma}$, where
${\langle\hat{O}\rangle_\sigma= \TrA\left(\hat{U}_\sigma\hat{O} \right)/\
\TrA\hat{U}_\sigma}$ is the thermal expectation value for $\hat{O}$ in the
canonical system with $\mathcal{A}$ valence nucleons for a given configuration of the $\sigma$ fields.

The multidimensional path integral in the Hubbard-Stratonovich representation can be calculated stochastically
using Monte Carlo techniques.   The thermal expectation value in Eq.~\eqref{CanDef} is estimated by
${\langle\hat{O}\rangle\approx \sum_k^N
\langle\hat{O}\rangle_{\sigma_k}\Phi_{\sigma_k}/\ \sum_k^N \Phi_{\sigma_k}}$,
which is obtained from a finite set of uncorrelated configurations of fields  $\sigma_k$ selected according to the distribution $W_\sigma$.

\ssec{Imaginary-time correlation matrices}   We denote the single-particle orbitals of the shell model space by $a \equiv (n_a, l_a, j_a)$. For each angular momentum $K$ and parity $\pi$, we define a density response  (or correlation) matrix  
\be\label{Cacbd}
C^{K \pi}_{ac,bd}(\tau) = \left\langle
\sum_M\rho^\dagger_{KM}(ac,\tau)\rho_{KM}(bd)\right\rangle\;,
\ee
where the angular-momentum coupled densities are defined by
${\rho_{KM}(ac)=\left[a^\dagger_a\times\tilde{a}_c\right]^{KM}}$ with ${\tilde{a}_{j_cm_c}=(-)^{j_c-m_c} a_{j_c\mn m_c}}$.
$\rho^\dagger_{KM}(ac,\tau)$ are the imaginary-time evolved densities given by ${\rho^\dagger_{KM}(ac,\tau) = e^{\tau\hat{H}}\rho_{KM}^\dagger(ac)e^{-\tau\hat{H}}}$. We will refer to the matrix $C^{K \pi}_{ac,bd}(\tau)$ as the imaginary-time correlation matrix (ITCM).

In the limit of large $\beta$, the thermal expectation value in (\ref{Cacbd}) reduces to an expectation value in the ground-state $|0\rangle$,  which has spin $0$ for even-even nuclei.  The ITCM assumes its spectral decomposition
\begin{equation}
C^K_{ac,bd}(\tau) = \sum_\alpha
v^*_{(ac)\alpha}v_{(bd)\alpha}
e^{-\tau\Delta E_{\alpha K}}\,,\label{CGrounProjTrun}
\end{equation}
where ${v_{(ac)\alpha}=(\alpha K||\rho_K(ac)||0)}$ are reduced one-body transition densities and $\Delta E_{\alpha
K}=E_{\alpha K}-E_0$ are excitation energies of states with angular momentum $K$. The sum over $\alpha$ in Eq.~\eqref{CGrounProjTrun} corresponds to a
specific parity $\pi$, and as a result, the ITCM carries a good parity. For simplicity of the notation, we omit here and in the following the parity label $\pi$. 

When $\tau$ is sufficiently large (but still $\tau \ll\beta$), or the excitation energies with angular momentum $K$ display a large gap, we can truncate the sum over $\alpha$ to $N_K$ terms, where $N_K$ is the dimension of the matrix $C^K(\tau)$
\begin{equation}
C^K_{ac,bd}(\tau) \approx \sum_{\alpha=1}^{N_K}
v^*_{(ac)\alpha}v_{(bd)\alpha}
e^{-\tau\Delta E_{\alpha K}}\;.
\end{equation}

In the following we assume the $N_K$ vectors  ${v_\alpha=(v_{(ac)\alpha})}$ (of dimension $N_K$) to be linearly
independent,  in which case they span a manifold of dimension $N_K$ and $C^K(\tau)$  is a positive-definite matrix.   Defining a biorthogonal basis $\psi_\beta$ in this manifold by $(\psi_\beta, v_\alpha) = \sum_{i=1}^N \psi_{i\beta} v_{i\alpha} = \delta_{\alpha,\beta}$, the vectors $\psi_\beta$ can be shown to be generalized eigenvectors of $C^K(\tau)$  with $C^K(\tau_0)$ as the weight function  
\begin{align}
C^K(\tau)\psi_\beta
&=\lambda_\beta(\tau-\tau_0) C^K(\tau_0)\psi_\beta\,,\label{GEVP}
\end{align}
and generalized eigenvalues of 
$\lambda_\beta(\tau-\tau_0)= e^{-\left|\tau-\tau_0\right|\Delta E_{\beta K}}$. Thus the many-body excitation energies $\Delta E_{\beta K}$ of states with angular momentum $K$ (and parity $\pi$) can be extracted from the generalized eigenvalue problem (GEVP) in Eq.~\eqref{GEVP}.

In practice, the vectors $v_\alpha$ are not linearly independent and the weight matrix $C(\tau_0)$ is positive semi-definite with several zero eigenvalues (within statistical errors), the precise number of which depends on $\tau_0$. In this cases the GEVP is
ill defined, and a reduction in the rank of the ITCM leads to a GEVP with generalized eigenvalues 
related to excitation energies as above.

The ITCM $C^K(\tau)$ can be cast in a form suitable for SMMC calculations. We define 
\be 
{\rho_{KM\sigma}(ac,\tau)=\left[\left({\bf
U}^{-\text{T}}_\sigma(\tau) a^\dagger\right)_i\times \left({\bf U}_\sigma(\tau)
\tilde{a}\right)_k\right]^{KM}} \;,
\ee
where ${i=(am_a)}$ and ${k=(cm_c)}$ label
single-particle states with good magnetic quantum number, ${{\bf
U}^{-\text{T}}_\sigma(\tau)=\left({\bf U}^{-1}_\sigma(\tau)\right)^\text{T}}$, and ${\bf U}_\sigma(\tau)$
describes the matrix representation of ${\hat{U}_\sigma(\tau)}$ in the
single-particle space. Then, the application of the Hubbard-Stratonovich
transformation to the imaginary-time evolved densities results in
\begin{align}
C^K_{ac,bd}(\tau)
&=\sum_M\overline{\left\langle \rho^\dagger_{KM\sigma}(ac,\tau)\rho_{KM}
\right\rangle_\sigma}
\,,\label{Csmmc4tmp}
\end{align}
where we have used the notation of Eq.~(\ref{CanDef}).\\

Above a certain value of $\tau$, it is necessary to stabilize the SMMC calculation of the ITCM. The method of stabilization for dynamical observables will be discussed elsewhere.

\ssec{Validation}
We next validate the ITCM method for an $sd$-shell nucleus $^{20}$Ne for which we can calculate the exact spectrum using the CI shell model code BIGSTICK~\cite{Johnson2018}.  The single-particle energies are taken from the USD interaction~\cite{Brown2006} and the interaction is an attractive quadrupole-quadrupole interaction $-\chi \tilde Q \cdot \tilde Q$, with $\tilde Q_{2\mu} = \sum_i r_i^2 Y_{2 \mu}(\hat r_i)$  and  $\chi = {8 \pi \over 5} {38.5\over A^{5/3}}$ MeV$/b^4$~\cite{Lauritzen1989}.

We illustrate the ITCM method  for $K^\pi=2^+$. 
The one-body densities $\rho_{KM}(ac)$ correspond to a specific
nucleon type, which leads to the ITCM block structure of the form $(\nu\nu\times
\nu'\nu')$, where $\nu$ denotes protons or neutrons.

We calculated the ITCM using both SMMC and BIGSTICK~\cite{Johnson2018}. The SMMC
calculations were carried out for ${\beta=10\text{ MeV}^{-1}}$ after extrapolating
to $\Delta\beta=0$ using the time-slices $\Delta\beta=$ 1/32 and 1/64 MeV$^{-1}$; in the case of BIGSTICK the calculations used the 
exact many-particle spectrum and one-body transition densities ${(\alpha J||\rho_K(ac)||\alpha'J')}$.

\begin{figure*}[ht!]
  \begin{center}
    \includegraphics[angle= 0,width=2\columnwidth]{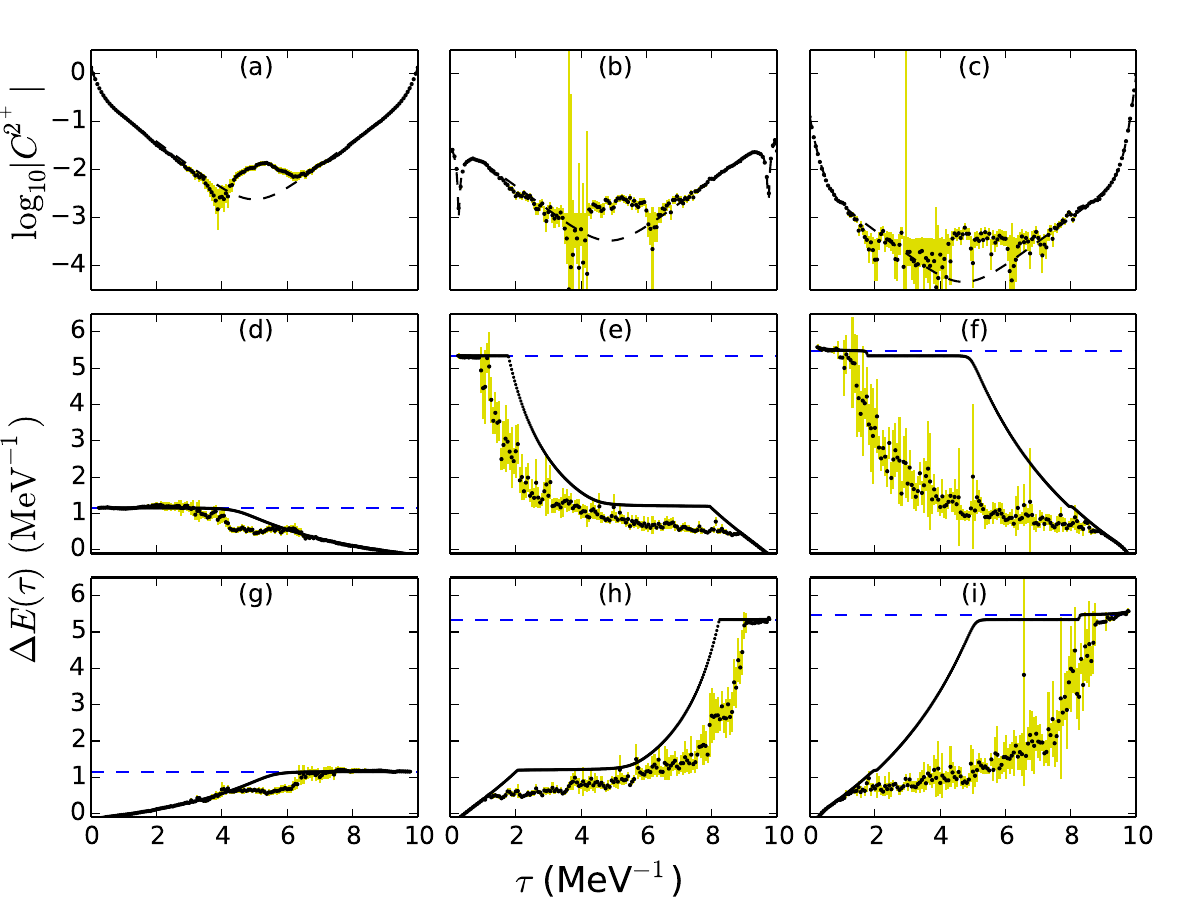}
    \caption{Panels (a)-(c) in top row: three representative
$K^\pi=2^+$ matrix elements at $\beta=10$ MeV$^{-1}$ as a function of $\tau$. The exact 
results obtained from BIGSTICK (dashed black lines) are compared to the SMMC 
calculations (solid circles with yellow error bars). Panels (d)-(f) in middle row: 
excitation energies $-\ln\lambda_\gamma/|\tau-\tau_0|$ vs.~$\tau$ for the first three $2^+$ levels  using the ITCM method with
$\beta=10\text{ MeV}^{-1}$ and $\tau_0=3/16\text{ MeV}^{-1}$. The ITCM estimates using BIGSTICK are plotted in black, and the corresponding SMMC 
results are shown in solid circles with yellow error bars. Panels (g)-(i) in bottom row: same as in panels (d)-(f) but using backward propagation in imaginary time. 
}\label{fig:CEx}
  \end{center}
\end{figure*}

In the top row of Fig.~\ref{fig:CEx} we show three representative ITCM elements for $K=2^+$.  The dashed black lines are the BIGSTICK calculations and the solid circles with yellow error bars are the SMMC calculations. The SMMC reproduces the exact results for small and large values of
$\tau$. The top panels of Fig.~\ref{fig:CEx} also illustrate the SMMC resolution
of $\sim 10^{-2}-10^{-3}$ for the matrix elements. As $|\tau-\tau_0|$ increases we leave the
asymptotic regime and the exponential factors $e^{-\Delta
E_\gamma(\beta-\tau)}$ systematically suppress the ITCM elements below the
SMMC's resolution, where statistical noise dominates; this suppression happens
as $|\tau|\to\beta/2$ and intensifies with increasing value of $\beta$. As can
be seen from the top panels of Fig. \ref{fig:CEx} the effect of the suppression
increases from left to right, which means that different matrix elements are
affected more than others. This effect is the SMMC
resolution and it is not due to numerical instability.

In practice, the matrix ${C^K(\tau_0)}$ can be non-positive definite because of statistical noise. When this happens the GEVP is
ill defined, and the ITCM cannot be used to estimate excitation energies.   In such a case we truncate the matrix ${C^K(\tau)}$  to the subspace for which the weight  matrix $C^K(\tau_0)$ is
positive definite; it can be shown that in the presence of zero eigenvalues, the
reduced ITCM still satisfies a GEVP with GEVs related to excitation energies as before.

In our example, the matrix $C^{2^+}(\tau_0)$ has dimension 16 with two zero eigenvalues (within SMMC resolution) for
$\beta=10\text{ MeV}^{-1}$, $\Delta\beta=1/64\text{ MeV}^{-1}$, and
$\tau_0=3/16$.  We truncate the dimension of the matrix $C^{2^+}(\tau)$ to 14 by restricting to the basis with positive eigenvalues for the weight matrix $C^K(\tau_0)$.

We find the eigenvalues $\lambda_\gamma$ of the truncated ITCM and calculate the ``excitation energies'' $-\ln\lambda_\gamma/|\tau-\tau_0|$ vs.~$\tau$. The panels (d-f) in the middle row of  Fig.~\ref{fig:CEx} show the results for the first three $2^+$ levels using a forward propagation.  The 
curves for $-\ln\lambda_\gamma/|\tau-\tau_0|$ calculated from the exact $C^{2}(\tau)$ matrix are shown in black, and the SMMC results are shown solid circles with yellow error bars.  The excitation energies calculated from direct diagonalization are shown by the blue constant dashed lines.
We observe that the SMMC results reproduces the exact excitation energies in a region where a plateau as a function of $\tau$.  This plateau  tends to be narrower for higher excitations. 

It is possible to extract excitation energies by a backward propagation in imaginary time. The bottom panels (g)-(i) in Fig.~\ref{fig:CEx}  show the corresponding results for the same three lowest $2^+$ levels using a backward propagation. 

The SMMC results become more susceptible to statistical noise with higher excitation energies. Contaminations
from higher excited states are more pronounced for higher excitation energies;
this is manifested in a systematic deviation for small $|\tau-\tau_0|$ observed
even in the exact BIGSTICK calculations (not shown). As a result only eight $2^+$ excitations are accessible in our example. 

Finite-$\beta$ effects are manifest in a monotonic decrease as $|\tau-\tau_0|$
increases as can be seen in Fig. \ref{fig:CEx}. The stair-case behavior is also
a finite-$\beta$ effect. There is a transition in the GEVs from the forward to
the backward asymptotic regime that results from the competition between $\tau$
and $\beta$. During this transition, the GEVs reorganize as previously
negligible terms in the spectral decomposition become more dominant. The
stair-case that results from this reorganization shortens the plateau and
restricts the range of $\tau$s which can be used to estimate excitation
energies. Higher excitation energies are affected the most.

\begin{figure}[ht!]
  \begin{center}
    \includegraphics[angle= 0,width=\columnwidth]{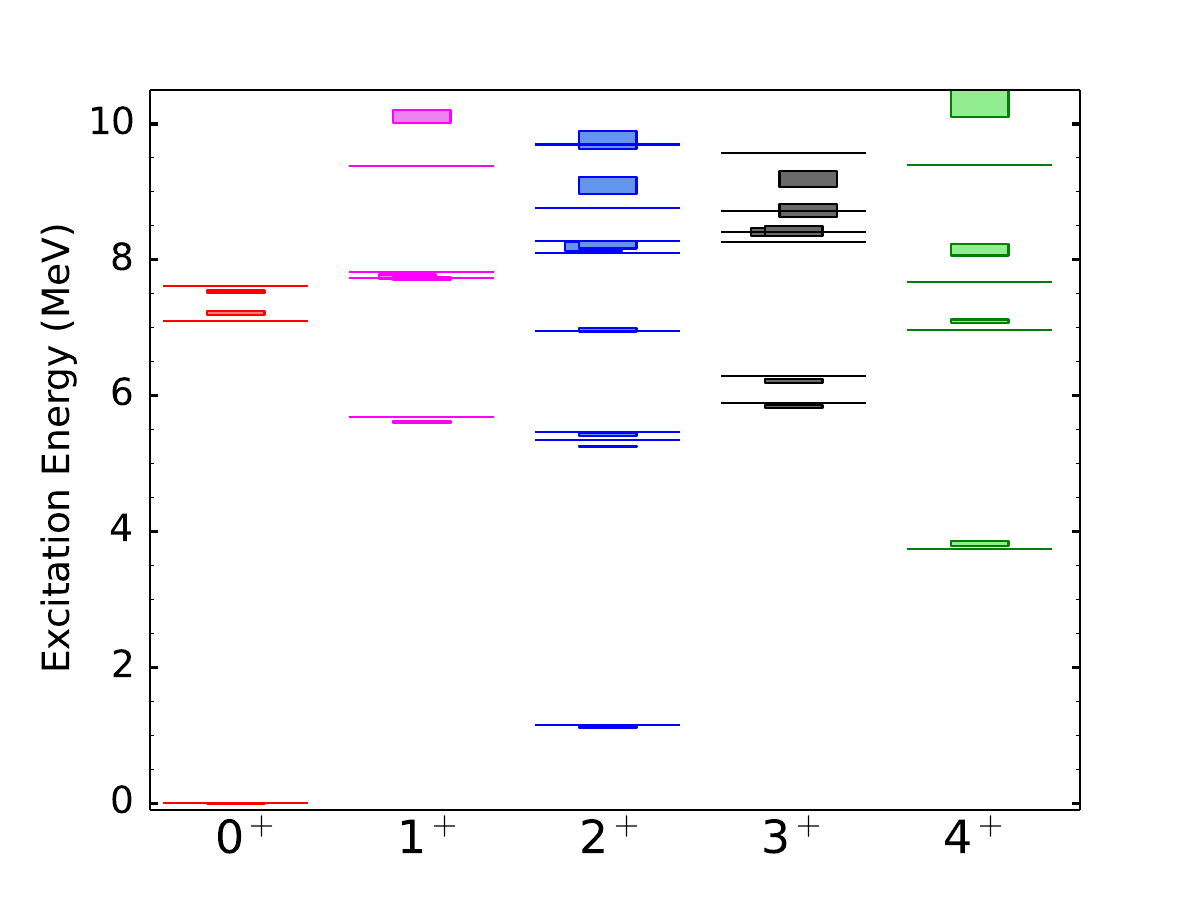}
    \caption{Excitation energies in $^{20}$Ne. The
solid lines are exact excitation energies obtained from diagonalizing the
CI shell model Hamiltonian using BIGSTICK. The SMMC estimates from the ITCM are shown with
statistical errors by the boxes centered around the average values.}
    \label{fig:Extab}
  \end{center}
\end{figure}

The allowed spin-parity combinations in the $sd$-shell model-space are $K^\pi=
0^+, 1^+, 2^+, 3^+$ and $4^+$. Results for the excitation energies
extracted from the SMMC ITCM for these $K^\pi$ are shown in
Fig.~\ref{fig:Extab}. The SMMC values are shown in the colored boxes and are
compared to exact excitation energies obtained from direct diagonalization
(solid horizontal lines).   
Table \ref{table:AllEx} summarizes the results where the number of levels shown extracted from SMMC 
 is less than the dimension of the corresponding truncated ITCM because statistical errors and contamination prohibit the extraction of higher excitation energies.
 With the exception of a few higher energy levels, the energy levels extracted in SMMC are in good agreement with their exact values.

\begin{table}[h!]
\centering
\begin{tabular}{lp{0.2cm} r@{.}l r@{.}l |p{0.3cm}p{0.3cm} r@{.}l r@{.}l}
\multicolumn{12}{c}{$E_\text{x}$ (MeV)}\\
\hline\hline
$K^+$&& \multicolumn{2}{c}{SMMC} & \multicolumn{2}{c}{Exact}&
$K^+$&& \multicolumn{2}{c}{SMMC} & \multicolumn{2}{c}{Exact}\\
\hline
\rule{0pt}{3ex}\multirow{3}{*}{$0^+$} 
  &&  0&00        &  0&0  &
\rule{0pt}{3ex}\multirow{6}{*}{$3^+$}
  &&  5&84(2)    & 5&9\\
  &&  7&22(3)    &  7&1  &&&  6&21(3)    & 6&3\\
  &&  7&53(2) &  7&6  &&& 8&41(6) & 8&3\\
\cline{1-6}
\rule{0pt}{3ex}\multirow{4}{*}{$1^+$}
 &&  5&61(1)  &  5&7 &
 && 8&42(7)  & 8&4\\
 &&  7&73(3)  &  7&7 &&& 8&73(10)  & 8&7\\
 &&  7&75(3) &  7&8 &&& 9&19(12)  & 9&6\\
 \cline{7-12}
 && 10&11(10) & 9&4 &&& 3&82(4) & 3&7\\
\cline{1-6}
\rule{0pt}{3ex}\multirow{8}{*}{$2^+$}
 && 1&12(1)      & 1&2  &
\rule{0pt}{3ex}\multirow{2}{*}{$4^+$}
 && 7&09(3)    & 7&0\\
 && 5&25(1)      & 5&3  &&& 8&15(9)   & 7&7\\
 && 5&43(2)      & 5&5  &&& 10&26(25) & 9&4\\
 && 6&97(3) & 7&0  &&& \mc{2}{c}{} & \mc{2}{c}{}\\
 && 8&19(6) & 8&1  &&& \mc{2}{c}{} & \mc{2}{c}{}\\
 && 8&22(5) & 8&3  &&& \mc{2}{c}{} & \mc{2}{c}{}\\
 && 9&09(13) & 8&8  &&& \mc{2}{c}{} & \mc{2}{c}{}\\
 && 9&76(13) & 9&7  &&& \mc{2}{c}{} & \mc{2}{c}{}\\
\hline
\hline
\end{tabular}
\caption[$^{20}$Ne excitation energies]{Exact low lying excitation energies for
$^{20}$Ne and their SMMC estimates from ITCM. The dimension of the ITCM (after truncation) exceeds the number of SMMC energies obtained due to effects described in the main text.
\label{table:AllEx}}
\end{table}

\ssec{Conclusion}   We have introduced and validated a systematic method within SMMC that employs imaginary-time correlation matrices (ITCM) to
calculate nuclear spectra. The ITCM satisfies a generalized eigenvalue problem (GEVP) for which the generalized eigenvalues are simply related to the excitation energies of low-lying states. We tested the ITCM method for the $sd$-shell nucleus $^{20}$Ne, and reproduced the excitation energies of low-lying  states calculated by exact diagonalization methods. The ITCM method can be applied within very large model spaces for which exact diagonalization methods are prohibited.  The method is applicable to other quantum many-body systems that can be described by a CI shell model approach. 

\ssec{Acknowledgements}  This work was supported in part by the U.S. DOE grant No.~DE-SC0019521. The calculations used resources of the National Energy Research Scientific Computing Center (NERSC), a U.S. Department of Energy Office of Science User Facility operated under Contract No.~DE-AC02-05CH11231.  We thank the Yale Center for Research Computing for guidance and use of the research computing infrastructure.


%

\end{document}